\documentclass[journal]{IEEEtran}
\usepackage{color}
\ifCLASSINFOpdf
\else
\fi

\usepackage{graphicx}      
\usepackage[square,sort,comma,numbers]{natbib}        
\usepackage{amssymb}
\usepackage{amsmath}
\usepackage{algorithm}
\usepackage{algorithmicx}
\usepackage{algpseudocode}
\usepackage{CJKutf8}

\newtheorem{defn}{Definition}
\newtheorem{rem}{Remark}
\newtheorem{thm}{Theorem}

\newtheorem{exmp}{Example}
\newtheorem{pf}{Proof}
\newtheorem{prop}{Proposition}

\begin{document}
%
\title{Reduced-Complexity Verification for $K$-Step and Infinite-Step Opacity in Discrete Event Systems}
%
%
%

\author{Xiaoyan Li,
        Christoforos N. Hadjicostis,~\IEEEmembership{Fellow,~IEEE,}
        and~Zhiwu Li,~\IEEEmembership{Fellow,~IEEE}
\thanks{X. Li is with the school of Information and Communication Engineering, North University of China, Taiyuan 030051, China (e-mail: 20220181@nuc.edu.cn)}
\thanks{C. N. Hadjicostis is with the Department of Electrical and Computer Engineering, University of Cyprus, Nicosia, Cyprus
(e-mail: hadjicostis.christoforos@ucy.ac.cy).}
\thanks{Z. Li is with the Institute of Systems Engineering, Macau University of Science and Technology, Macau SAR, China, and with the School of Mechano-Electronic Engineering, Xidian University, Xi'an 710071, China (e-mail: zhwli@xidian.edu.cn).}}

\maketitle

\begin{abstract}
Opacity is a property that captures security concerns in cyber-physical systems and its verification plays a significant role.
This paper investigates the verifications of $K$-step and infinite-step weak and strong opacity for partially observed nondeterministic finite state automata.
$K$-step weak opacity is checked by constructing, for some states in the observer, appropriate state-trees, to propose a necessary and sufficient condition.
Based on the relation between $K$-step weak and infinite-step weak opacity, a condition that determines when a system is not infinite-step weak opaque is presented.
Regarding $K$-step and infinite-step strong opacity, we develop a secret-involved projected automaton, based on which we construct secret-unvisited state-trees to derive a necessary and sufficient condition for $K$-step strong opacity.
Furthermore, an algorithm is reported to compute a verifier that can be used to obtain a necessary and sufficient condition for infinite-step strong opacity.
It is argued that, in some particular cases, the proposed methods achieve reduced complexity compared with the state of the art.
\end{abstract}


%
\IEEEpeerreviewmaketitle

\section{Introduction}\label{Introduction}
Opacity requires that during the evolution of a system, the secrets of a partially-observable system should not be exposed to an outside intruder that has perfect knowledge of the system transition structure but only perceives the occurrence of observable events (and can thus estimate the system possible next states, using available information about the possible current states and the observed events at hand) \cite{Hadjicostis:2020}, \cite{Lafortune:2018}, \cite{Jacob:2016}, \cite{guo2020overview}, \cite{HAN2023110756}, \cite{YIN2019266}.
Current-state opacity, language-based opacity, initial-state opacity, initial-final state opacity, $K$-step opacity, and infinite-step opacity have been presented to capture different scenarios of secrets that should not be revealed, and corresponding verification approaches for checking these notions have been developed \cite{Lin:2011}, \cite{Saboori:2011}, \cite{Saboori:2012}, \cite{Saboori:2013}, \cite{Falcone:2015}, \cite{Xiang:2017}, \cite{Balun:2021}, \cite{Ma:2021}, \cite{2022Verification}, \cite{Wintenberg:2022}.
When a system violates the desired opacity property, dynamic mask \cite{Xiang:2020} and obfuscation mechanisms, including insertion function \cite{wu2014synthesis,liu2022enforcement}, embedded insertion function \cite{2021Embedded}, extended insertion function \cite{li2021extended}, and edit function \cite{2017Enforcing} are proposed to enforce opacity.

In \cite{Xiang:2017}, the authors construct a two-way observer to verify infinite-step weak opacity with complexity of $O(|E_o|2^{2|X|})$, and develop an algorithm for checking $K$-step weak opacity with complexity of $O(\min\{|E_o|2^{2|X|}, |E_o|^{K+1}2^{|X|}\})$, where $X$ and $E_o$ are the sets of states and observable events in the considered automaton, respectively.
In \cite{Balun:2021}, the authors indicate that $K$-step weak opacity and infinite-step weak opacity can also be transformed to current-state opacity, language-based opacity, initial-state opacity, and initial-final state opacity, based on which, approaches to verify $K$-step weak opacity and infinite-step weak opacity are presented; the corresponding computational complexities are $O((K+1)2^{|X|}(|X|+m|E_o|^2))$ and $O((|X|+m|E_o|)2^{|X|})$, where $m\leq |E_o||X|^2$ is the number of transitions in the so-called projected automaton \cite{Hop:1979} of the considered automaton.
In \cite{Ma:2021}, an infinite-step recognizer is proposed, which is an information structure that can be used to check infinite-step strong opacity; the complexity of the corresponding verification method is $O(2^{2|X|}|E_o|)$. Furthermore, a $K$-step recognizer is constructed to check $K$-step strong opacity for a given value of $K$, with complexity of $O(2^{(K+2)|X|}|E_o|)$.

In \cite{Wintenberg:2022}, in order to unify many existing notions of opacity for discrete event systems, a general framework is presented.
Based on the framework, the authors also investigate verification methods for $K$-step weak and strong opacity and infinite-step weak opacity.
The state complexities to check $K$-step weak opacity are $2^{|X|(K+3)}$ using the secret observer and $|X|(K+1)3^{|X|}$ using the reverse comparison approach, respectively.
In \cite{Wintenberg:2022}, the state complexities to check $K$-step strong opacity are $(K+3)^{|X|}$ using the secret observer and $K2^{|X|}$ using the reverse comparison approach, respectively.
The state complexity of verifying infinite-step weak opacity is the same as the method in \cite{Xiang:2017} and is $2^{2|X|}$.

This paper focuses on the verification of $K$-step weak and strong opacity, and infinite-step strong opacity.
In order to verify $K$-step weak opacity, state-trees for some states in the observer of the considered system are constructed and used to derive a necessary and sufficient condition.
For $K$-step and infinite-step strong opacity, a secret-involved projected automaton is proposed, based on which we construct secret-unvisited state-trees to derive a necessary and sufficient condition for $K$-step strong opacity, and develop an algorithm to compute a verifier that provides a necessary and sufficient condition for infinite-step strong opacity.
The complexity of verifying $K$-step weak and strong opacity is $O(\max\{|E_o|^{K+1}(2^{|X|}-2^{|X_{NS}|}-2^{|X_S|}),2^{|X|}|E_o|\})$ with $X_{NS}$ and $X_{S}$ being the sets of non-secret and secret states.
As described later in the paper, our strategies have, in some cases, significant complexity advantages compared with the existing approaches in \cite{Balun:2021} and \cite{Ma:2021};
in particular, the complexity of verifying infinite-step strong opacity is $O(2^{2|X|}|E_{o}|)$, which is the same as that in \cite{Ma:2021}; however, our checking method is also applicable to nondeterministic finite automata.

\section{Preliminaries}\label{Preliminaries}
We use $E^*$ to denote the set of all finite-length strings of elements of $E$, including the empty string $\varepsilon$.
Given two strings $s_1, s_2\in E^{*}$, their concatenation is denoted by $s_1s_2$, implying that the sequence of events captured by $s_1$ is immediately followed by that captured by $s_2$.
For a string $s\in E^{*}$, $s'\in E^{*}$ is called a prefix of $s$ if there exists $s''\in E^{*}$ such that $s=s's''$, and the set of all prefixes of $s$ is denoted by $\overline{s}$.
For a string $s\in E^{*}$, $s''\in E^{*}$ is called a suffix of $s$ if there exists $s'\in E^{*}$ such that $s=s's''$, which is denoted by $s''=s/s'$.
The length of a string $s$ is denoted by $|s|$. Note that for the empty string $\varepsilon$, we have $|\varepsilon|=0$.
We use $s[i]$ to denote the $i$-th event in a nonempty string $s$, where $i\in \{1, 2, \ldots, |s|\}$.
For two sets $A$ and $B$, $p\in A\setminus B$ denotes that $p\in A$ and $p\notin B$. We use $A[i]$ to denote the $(i+1)$-st element in a finite $A$ (under an arbitrary ordering), where $i\in \{0,1,...,|A|-1\}$ and $|A|$ is the cardinality of $A$.


A nondeterministic finite automaton (NFA) is a four-tuple ${G_{nd}=(X, E, \delta, X_0)}$, where $X$ and $E$ are the sets of states and events, respectively, $\delta: X \times E \rightarrow 2^X$ is the nondeterministic transition function, and $X_0 \subseteq X$ is the set of possible initial states. (When $|X_0|=1$ holds and $\delta: X \times E \rightarrow 2^X$ reduces to a possibly partially defined $f: X \times E \rightarrow X$, an NFA reduces to a deterministic finite automaton (DFA).)
The transition function $\delta$ can be extended from $X \times E \rightarrow 2^X$ to $X \times E^{*} \rightarrow 2^X$ in a recursive way, which is defined as $\delta(x, es)=\delta(\delta(x,e),s)=\bigcup_{x'\in\delta(x, e)}\delta(x', s)$, where $e\in E$ and $s\in E^{*}$ (specifically, $\delta(x, \varepsilon)=\{x\}$ for all $x\in X$).
Note also that $\delta(X,s)=\bigcup_{x\in X}\delta(x,s)$.
The behavior of a given NFA is captured by its generated language, which is defined as $L(G_{nd})=\{s\in E^{*}|(\exists x_0\in X_0)  [\delta(x_0, s)\neq \emptyset]\}$. We also define the system behavior starting from a state $x\in X$ as $L(G_{nd}, x)=\{s\in E^{*}|\delta(x, s)\neq \emptyset$\}.

Typically, an automaton is assumed to be partially observable, which indicates that the events of the system $E$ can be partitioned into two disjoint subsets: the set of observable events $E_o$ and the set of unobservable events $E_{uo}$, i.e., $E= E_o \dot{\cup}E_{uo}$.
The natural projection of an event is defined as $P(e)=e$ if $e\in E_o$; otherwise, $P(e)=\varepsilon$ if $e\in E_{uo}$.
Given a string $s\in E^{*}$, its observation is the output of the natural projection $P: E^{*}\rightarrow E_o^{*}$, which is defined recursively as $P(es')=P(e)P(s')$, where $e\in E$ and $s'\in E^{*}$.

Given an NFA $G_{nd}$, for a state $x$, we use
$T_{o}(x)=\{e_{o}\in E_{o}|(\exists x'\in X\wedge s\in E^*)[x'\in \delta(x,s)\wedge P(s)=e_o]\}$ to denote the set of events that can be observed at $x$,
and $UR(x)=\{x'|(\exists s\in E^*) [x'\in \delta(x,s)\wedge P(s)=\varepsilon]\}$ to denote the unobservable reach at state $x$.
For a state set $X_1$, $UR(X_1)=\bigcup_{x\in X_1}UR(x)$ denotes the unobservable reach at state set $X_1$, and
$R(X_1,e_{o})=\{x\in X|(\exists x'\in X_1,\exists s\in E^{*})[x\in \delta(x',s)\wedge P(s)=e_{o}]\}$ denotes the observable reach following an observable event $e_{o}$ at $X_1$.
For an automaton, we use $(x,e,x')$ to denote a transition, implying that state $x'$ can be reached from state $x$ via the event $e$.

\section{Opacity Notions}\label{Opacity-Definitions}
This section describes the opacity notions involved in this paper, including $K$-step weak and strong opacity and infinite-step weak and strong opacity.
For an NFA, the set of secret states and the set of non-secret states are denoted by $X_{S}$ and $X_{NS}$, respectively. Typically, $X=X_{S} \cup X_{NS}$ and $X_{S} \cap X_{NS}=\emptyset$ hold, i.e., $X_{NS}=X \setminus X_{S}$.
Intuitively, an NFA is $K$-step weak opaque, if for all possible initial states $x_{0}\in X_{0}$ and for all strings $st\in L(G_{nd},x_{0})$ reaching a secret state after $s$ (i.e., $\delta(x_{0},s)\cap X_{S}\neq \emptyset$ and $\delta(\delta(x_{0},s)\cap X_{S},t)\neq \emptyset$) with $|P(t)|\leq K$, there exist an initial state $x_{0}'$ and an observationally equivalent string from this initial state $s't'\in L(G_{nd},x_{0}')$ (specifically, $P(s)=P(s')$ and $P(t)=P(t')$) such that the NFA reaches a non-secret state from $x_{0}'$ after $s'$ (i.e., $\delta(x_{0}',s')\cap X_{NS}\neq \emptyset$ and $\delta(\delta(x_{0}',s')\cap X_{NS},t')\neq \emptyset$).
This is formulated in Definition~\ref{$K$-step-weak-opacity}.

\begin{defn}\cite{Saboori:2011}\label{$K$-step-weak-opacity}
Given an NFA $G_{nd}=(X,E,\delta,X_0)$ with set of secret states $X_S\subseteq X$, set of observable events $E_{o}\subseteq E$, and natural projection $P$, $G_{nd}$ is said to be $K$-step weak opaque for $K\geq 0$ if
$$(\forall x_{0}\in X_{0},\forall st\in L(G_{nd},x_{0}))$$ $$[\delta(x_0,s)\cap X_S\neq \emptyset {\wedge \delta(\delta(x_{0},s)\cap X_{S},t)\neq \emptyset} \wedge |P(t)|\leq K)$$ $$\implies$$
$$(\exists x_{0}'\in X_{0},\exists s't'\in L(G_{nd},x_{0}'))$$
$$\delta(x_0',s')\cap X_{NS}\neq \emptyset {\wedge \delta(\delta(x_{0}',s')\cap X_{NS},t')\neq \emptyset}$$ $$\wedge P(s')=P(s)\wedge P(t')=P(t)].$$
\end{defn}

\begin{figure}[!htbp]
  \centering
  \includegraphics[width=0.9\linewidth]{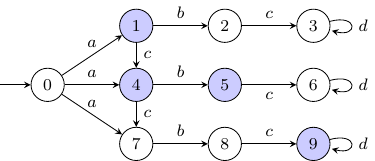}\\
  \caption{Example for $K$-step weak opacity.}\label{running-example-2}
\end{figure}

\begin{exmp}
We use an example to elucidate $K$-step weak opacity in Definition~\ref{$K$-step-weak-opacity}.
Consider the NFA in Fig.~\ref{running-example-2}, where the set of initial states is $X_{0}=\{0\}$, the set of observable events is $E_{o}=\{a,b,c,d\}$, the set of secret states is $X_{S}=\{1,4,5,9\}$ (the shaded states), and the set of non-secret states is $X_{NS}=\{0,2,3,6,7,8\}$.
Suppose that $K=1$. For string $ac$ with $P(ac)=ac$ and $|c|=1$, we have $\delta(0,a)\cap X_{S}=\{1,4\}\neq \emptyset$ and $\delta(\delta(0,a)\cap X_{S},c)=\{4,7\}\neq \emptyset$.
However, there does not exist another string $s't'$ from initial state 0 with $P(s')=a$ and $P(t')=c$  such that $\delta(0,s')\cap X_{NS}\neq \emptyset$ and $\delta(\delta(0,s')\cap X_{NS},t')\neq \emptyset$ (due to $\delta(0,a)\cap X_{NS}=\{7\}$ and $\delta(\{7\},c)=\emptyset$).
As a consequence, the NFA is not $1$-step weak opaque since when sequence of observations $ac$ is observed by the outside intruder, it can deduce with certain that the NFA is in a secret state (1 or 4) one step ago.
\hfill $\square$
\end{exmp}

We next introduce $K$-step strong opacity, which is strong version of $K$-step weak opacity. Note that the definition of $K$-step strong opacity below is identical with the notion of trajectory-based $K$-step opacity in \cite{Saboori:2011} and is described in more detail in Section~III.B. More specifically, $K$-step strong opacity requires that for each observation sequence that a system outputs, there exists a trajectory such that (i) it is observationally equivalent to the observation sequence, and (ii) there exists at least one sequence of states following the trajectory such that no secret state is visited over the last $K$-steps.
The formal definition is as follows.

\begin{defn}\label{$K$-step-strong-opacity}
Given an NFA $G_{nd}=(X,E,\delta,X_0)$ with set of secret states $X_S\subseteq X$, set of observable events $E_{o}\subseteq E$, and natural projection $P$, $G_{nd}$ is said to be $K$-step strong opaque for $K\geq 0$ if
$$(\forall x_{0}\in X_{0}, \forall st\in L(G_{nd},x_{0}))$$
$$[\delta(x_0,s)\cap X_S\neq \emptyset\wedge \delta(\delta(x_{0},s)\cap X_{S},t)\neq \emptyset \wedge |P(t)|\leq K$$
$$\implies$$
$$(\exists x_{0}'\in X_{0}, \exists \omega\in L(G_{nd},x_{0}'), \forall \omega'\in \bar{\omega})$$
 $$[|P(\omega)|-|P(\omega')|\leq K\implies {\delta(x_0',\omega')\cap X_{NS}= X^{1}\neq \emptyset} \wedge$$ $$(\forall i \in \{1,2,\ldots, |\omega|-|\omega'|\}) [j=|\omega'|+i\wedge \delta(X^{i},\omega[j])\cap X_{NS}\neq \emptyset$$ $$\wedge \delta(X^{i},\omega[j+1])=X^{i+1}\wedge X^{i+1}\cap X_{NS}\neq \emptyset]\wedge P(st)=P(\omega)]].$$
\end{defn}

In Definition~\ref{$K$-step-strong-opacity}, $\delta(x_0',\omega')\cap X_{NS}= X^{1}\neq \emptyset \wedge(\forall i \in \{1,2,\ldots, |\omega|-|\omega'|\}) [j=|\omega'|+i\wedge \delta(X^{i},\omega[j])\cap X_{NS}\neq \emptyset\wedge \delta(X^{i},\omega[j+1])=X^{i+1}\wedge X^{i+1}\cap X_{NS}\neq \emptyset]$ indicates that there necessarily exists a sequence of states $x_0'x_{1}x_{2}\cdots x_{|\omega|}$ following sequence of observations $\omega$ such that $x_{i}\in X^{i}\cap X_{NS}$ for all $i\in \{|\omega'|,|\omega'|+1,\ldots,|\omega|\}$ and $x_{j+1}\in \delta(\{x_{j}\},\omega[j+1])$ for all $i\in \{|\omega'|,|\omega'|+1,\ldots,|\omega|-1\}$, i.e., we have $\delta(\cdots\delta(\delta(\delta(x_0,\omega')\cap X_{NS},\omega/\omega^\prime[1])\cap X_{NS},\omega/\omega^\prime[2])\cap X_{NS} \cdots,\omega/\omega^\prime[|\omega/\omega^\prime|])\cap X_{NS}=\emptyset$.

\begin{figure}[!htbp]
  \centering
  \includegraphics[width=0.9\linewidth]{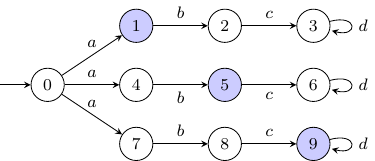}\\
  \caption{Example for $K$-step strong opacity and infinite-step weak opacity.}\label{running-example}
\end{figure}

\begin{exmp}\label{eg-1}
We use this example to illustrate the notion of $K$-step strong opacity in Definition~\ref{$K$-step-strong-opacity} and also the difference between $K$-step weak and strong opacity.
Consider the NFA in Fig.~\ref{running-example}, where the set of observable events is $E_{o}=\{a,b,c,d\}$, the set of secret states is $X_{S}=\{1,5,9\}$ (the shaded states), and the set of non-secret states is $X_{NS}=\{0,2,3,4,6,7,8\}$.
We use string $abc$ when $K=1$ as an instance. For this string, let $s=ab$ and $t=c$, and the following conditions hold: $\delta(x_0,ab)\cap X_S=\{5\}\neq \emptyset$, $\delta(\{5\},c)=\{6\}\neq \emptyset$ and $|P(t)|=1\leq K=1$.
There exists a string $w=abc$ that satisfies the following condition: for all $\omega'\in \bar{\omega}$ with $|P(\omega)|-|P(\omega')|\leq K=1$ (i.e., $\omega'\in \{ab,abc\}$), we have $\delta(0,ab)\cap X_{NS}=\{2,8\}\neq \emptyset$, $\delta(\{2,8\},c)=\{3\}\neq \emptyset$, and $P(st)=P(\omega)=abc$.
For other strings that the NFA may generate, the above descriptions apply.
Consequently, the NFA is $1$-step strong opaque, which is also implied by the followings: (i) $(1,b,2)$ and $(4,b,5)$ are confused by $(7,b,8)$, (ii) $(5,c,6)$ and $(8,c,9)$ are confused by $(2,c,3)$, and (ii) $(9,d,9)$ is confused by $(3,d,3)$ and $(6,d,6)$.

Suppose that $K=2$.
All possible state sequences following string $abc$ are listed as 0123, 0456, and 0789.
When the intruder observes $abc$, it can definitely infer that one of the secret states 1, 5, or 9 has been visited when it tracks back 2 steps ago, resulting that the NFA is not 2-step strong opaque. In other words, there does not exist a string $abc$ with $\delta(\delta(\delta(0,a)\cap X_{NS},b)\cap X_{NS},c)\cap X_{NS}\neq \emptyset$.
However, the NFA is 2-step weak opaque since when the intruder respectively observes (i) $ab$, (ii) $abc$, (iii) $abcd$, and (iv) $abcddd^*$, the NFA might be respectively in non-secret state (i) 0, (ii) 4 and 7, (iii) 2 and 8, and (iv) 3 and 6 when it tracks back two steps ago.
\hfill $\square$
\end{exmp}

\begin{rem}
As Example~\ref{eg-1} shows, the key difference between $K$-step weak opacity and $K$-step strong opacity is that when an intruder observes a sequence of observable events and tracks back to find if some secrets are revealed, the former focuses on the instant but the latter focuses on the period.
Note that $K$-step strong opacity in this paper is also the extension of $K$-step strong opacity formulated in \cite{Falcone:2015} from DFAs to NFAs since a trajectory corresponds to a unique sequence of states in terms of DFAs and a trajectory corresponds to multiple sequence of states when it comes to NFAs.
\end{rem}

When $K=0$, $K$-step weak (strong) opacity reduces to current-state opacity.
When $K\rightarrow \infty$, $K$-step weak (strong) opacity turns out to be infinite-step weak (strong) opacity. In what follows, we define infinite-step weak and strong opacity.

\begin{defn}\cite{Saboori:2012}\label{infinite-step-weak-opacity}
Given an NFA $G_{nd}=(X,E,\delta,X_0)$ with the set of secret states $X_S\subseteq X$, the set of observable events $E_{o}\subseteq E$, and the natural projection $P$, $G_{nd}$ is said to be infinite-step weak opaque if
$$(\forall x_{0}\in X_{0}, \forall st\in L(G_{nd},x_{0}))$$
$$[\delta(x_0,s)\cap X_S\neq \emptyset {\wedge \delta(\delta(x_{0},s)\cap X_{S},t)\neq \emptyset}\implies$$
$$(\exists x_{0}'\in X_{0},\exists s't'\in L(G_{nd},x_{0}'))$$ $$\delta(x_0',s')\cap X_{NS}\neq \emptyset {\wedge \delta(\delta(x_{0}',s')\cap X_{NS},t')\neq \emptyset}$$ $$\wedge P(s')=P(s)\wedge P(t')=P(t)].$$
\end{defn}

In Definition~\ref{infinite-step-weak-opacity}, infinite-step weak opacity requires that for every string reaching a secret state, there should exist another observationally equivalent string that reaches a non-secret state after having the same arbitrarily long sequence of observable events. In other words, the intruder cannot definitely determine that the system is/was in secret states based on the  sequence of observations observed thus far and any possible continuations of sequences of observations of arbitrary length.

\begin{exmp}
Reconsider the NFA in Fig.~\ref{running-example}.
The NFA is infinite-step weak opaque since when an intruder observes sequence $abcd^*$ and tracks any steps back, it cannot be certain that the system state is in secret state 1 or non-secret states 4 and 7, secret state 5 or non-secret states 2 and 8, and secret state 9 or non-secret states 3 and 6.
\hfill $\square$
\end{exmp}

\begin{defn}\label{infinite-step-strong-opacity}
Given an NFA $G_{nd}=(X,E,\delta,X_0)$ with set of secret states $X_S\subseteq X$, set of observable events $E_{o}\subseteq E$, and natural projection $P$, $G_{nd}$ is said to be infinite-step strong opaque if
$$(\forall x_{0}\in X_{0}, \forall st\in L(G_{nd},x_{0}))$$
$$[\delta(x_0,s)\cap X_S\neq \emptyset {\wedge \delta(\delta(x_{0},s)\cap X_{S},t)\neq \emptyset}\implies$$
$$(\exists x_{0}'\in X_{0}\cap X_{NS}, \exists \omega\in L(G_{nd},x_{0}'))$$  $$ \delta(x_{0}',\omega[1])=X^{1}\wedge X^{1}\cap X_{NS}\neq \emptyset
\wedge(\forall i \in \{1,2,\ldots, |\omega|-1\})$$
$$ \delta(X^{i},\omega[i+1])=X^{i+1}\wedge X^{i+1}\cap X_{NS}\neq \emptyset
\wedge P(st)=P(\omega)].$$
\end{defn}

\begin{figure}[!htbp]
  \centering
  \includegraphics[width=0.9\linewidth]{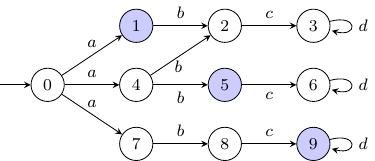}\\
  \caption{Example for infinite-step strong opacity.}\label{running-example-1}
\end{figure}

\begin{exmp}
This example is used to show the notion of infinite-step strong opacity. Consider the NFA in Fig.~\ref{running-example-1}, where the set of observable events is $E_{o}=\{a,b,c,d\}$, the set of secret states is $X_{S}=\{1,5,9\}$ (the shaded states), and the set of non-secret states is $X_{NS}=\{0,2,3,4,6,7,8\}$.
For string $st=abcd^*$ with $s=a$ and $t=bcd^*$, we have $x_{0}=0\in X_{0}=\{0\}$,  $\delta(0,a)=\{1,4,7\}$, $\{1,4,7\}\cap X_{S}=\{1\}\neq \emptyset$, and $\delta(\{1\},bcd^*)\neq \emptyset$. There exists $\omega=abcd^*\in L(G_{nd},0)$ such that $\delta(0,a)\cap X_{NS}=\{4,7\}$, $\delta(\{4,7\},b)\cap X_{NS}=\{2\}$, $\delta(\{2\},c)\cap X_{NS}=\{3\}$, and $\delta(\{3\},d^*)\cap X_{NS}=\{3\}\neq \emptyset$.
For all strings $st$ that can be generated by the NFA with $\delta(0,s)\cap X_S\neq \emptyset$ and $\delta(\delta(x_{0},s)\cap X_{S},t)\neq \emptyset$, $\omega=abcd^*\in L(G_{nd},0)$ satisfying $\delta(\cdots\delta(\delta(x_{0}',\omega[1])\cap X_{NS},\omega[2])\cap X_{NS}\cdots, \omega[|\omega|])\cap X_{NS}\neq \emptyset$ and $P(st)=P(\omega)$ can be found, implying that the NFA is infinite-step strong opaque (which is also illustrated by the fact that there exists a sequence of states $0423^*$ that does not contain any secret states corresponding to string $abcd^*\in L(G_{nd},0)$).
\hfill $\square$
\end{exmp}

\begin{rem}
In plain words, an NFA is said to be infinite-step strong opaque if (i) it is infinite-step weak opaque, and (ii) for each actual execution, there exists a trajectory that is observationally equivalent with the actual execution requiring that there exists at least one sequence of states following the trajectory such that no secret state is visited over any previous steps.
\end{rem}

%
%
%
%
\begin{defn}\cite{Hadjicostis:2020}\label{observer}
Consider an NFA $G_{nd}=(X,E,\delta,X_0)$, with respect to a set of observable events $E_{o}\subseteq E$ and natural projection $P$. The observer is given by a fully observable DFA  $G_{obs}=(X_{obs},E_{o},f_{obs},x_{obs,0})$, where the set of states is $X_{obs}\subseteq 2^X$, the set of events is $E_{o}$ and all events are observable, the initial state is defined as $x_{obs,0}=UR(X_{0})$, and
the transition function is defined as follows: for all $x_{obs}\in X_{obs}$ and $e_{o}\in E_{o}$, $f_{obs}(x_{obs},e_{o})$ = $R(x_{obs},e_{0})$.
\end{defn}

{\section{Verification for $K$-Step Weak Opacity}
This section investigates the verification method for $K$-step weak opacity by considering each state $x_{obs}\in X_{obs}$ with $x_{obs}\cap X_{S}\neq \emptyset$ as a root node and constructing a state-tree based on what follows.

Step~1: Find the root node by defining $rn=(x_1,x_2)$ with $x_{1}=\{x|x\in x_{obs}\cap X_{S}\}$ and $x_{2}=\{x|x\in x_{obs}\cap X_{NS}\}$.

Step~2: Based on the observer, for all $e_{o}\in E_{o}$ with $x_{obs}'=f_{obs}(x_{obs},e_{o})$, the immediate subsequent node corresponding to $x_{obs}'$ is obtained as $rn'=(x_1',x_2')$ with $x_{1}'=\{x\in x_{obs}'|(\exists x'\in x_1)[x\in R(\{x'\},e_o)]\}$ and $x_{2}'=\{x\in x_{obs}'|(\exists x'\in x_2)[x\in R(\{x'\},e_o)]\}$, i.e., the transition $(rn,e_o,rn')$ is added to the state-tree.

Step~3: Let $rn=rn'$ and repeat Step 2 $(K-1)$ times.

\begin{exmp}\label{eg-k-step-weak-opaque-yes}
Consider the DFA in Fig.~\ref{eg4}(a), with the set of observable events $E_{o}=\{a,b,c\}$, the set of unobservable events $E_{uo}=\{d\}$, and the set of secret states
(graphically represented by shaded circles) $X_S=\{2,7\}$. The observer is constructed as depicted in Fig.~\ref{eg4}(b) according to Definition~\ref{observer}. Suppose that we would like to verify 2-step weak opacity. Two state-trees need to be constructed with $\{2,5\}$ and $\{4,7\}$ being the considered root nodes, respectively, due to $X_S=\{2,7\}$.
We use $\{2,5\}$ as an instance to show the process of constructing a state-tree.
The root node is $(\{2\},\{5\})$ based on Step~1. Following Step~2, the transition $((\{2\},\{5\}),b,(\{3\},\{6\}))$ is added. For the new node (\{3\},\{6\}), we obtain the transition $((\{3\},\{6\}),a,(\{4\},\{7\}))$.
The state-tree corresponding to $\{2,5\}$ is shown in the left of Fig.~\ref{eg4}(c).
 $\hfill\square$
\end{exmp}

\begin{figure*}[!htbp]
  \centering
  \includegraphics[width=0.85\linewidth]{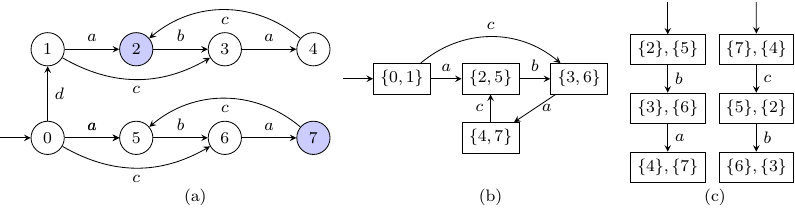}\\
  \caption{Example for constructing state-trees to verify 2-step weak opacity: (a) DFA with $X_{S}=\{2,7\}$; (b) Observer; (c) State-tree associated with $\{2,5\}$ and $\{4,7\}$.}\label{eg4}
\end{figure*}

\begin{thm}\label{$K$-step-weak-opaque-verification}
Consider an NFA $G_{nd}=(X,E,\delta,X_0)$ with respect to set of secret states $X_S\subseteq X$, set of observable events $E_{o}\subseteq E$, and natural projection $P$, and its observer $G_{obs}=(X_{obs},E_o,f_{obs},x_{obs,0})$. Construct all state-trees for all $x_{obs}\in X_{obs}$ such that $x_{obs}\cap X_S\neq \emptyset$. Let $X_{st}$ denote the set of all states in the obtained state-trees. $G_{nd}$ is $K$-step weak opaque if and only if $$(\forall x_{st}=(x_1,x_2)\in X_{st})x_2\neq \emptyset.$$
\end{thm}

\begin{pf}
\rm
($\Longleftarrow$) Suppose that $G_{nd}$ is not $K$-step weak opaque.
Our goal is to show that there exists an $x_{st}=(x_1,x_2)\in X_{st}$ such that $x_2=\emptyset$.
Since $G_{nd}$ is not $K$-step weak opaque, there necessarily exist a state $x_{0}\in X_{0}$ and a string $st\in L(G_{nd},x_{0})$ such that $\delta(x_0,s)\cap X_{S}\neq \emptyset$, $\delta(\delta(x_0,s)\cap X_{S},t)\neq \emptyset$, and $|P(t)|\leq K$, and for this string $st$, there do not exist a state $x_{0}'$ and a string $s't'\in L(G_{nd},x_{0}')$ satisfying $P(s')=P(s)$, $P(t)=P(t')$, $\delta(x_0',s')\cap X_{NS}\neq \emptyset$, and $\delta(\delta(x_0',s')\cap X_{NS},t')\neq \emptyset$.

Let $x\in \delta(x_0,s)\cap X_S$ with $\delta(x,t)\neq \emptyset$ and let $|P(t)|=n\leq K$.
Consider a state-tree corresponding to $x_{obs}=f_{obs}(x_{obs,0},P(s))$ with $x\in x_{obs}$.
We use $x_{st,0}x_{st,1}\cdots x_{st,n-1}x_{st,n}$ to denote a path in the state-tree.
For this path, due to the constructions of the observer and the state-tree, the following two conditions hold: (i) there exists a unique sequence of observations $e_{o,1}e_{o,2}\cdots e_{o,n}$ such that for all $i\in \{1,2,\ldots,n\}$, we have the transition $(x_{st,i-1},e_{oi},x_{st,i})$;
(ii) we have $x_{st,0}=\{x_{1,0},x_{2,0}\}$ with $x_{1,0}=\{x'|x'\in X_{S}\cap x_{obs}\}$ and $x_{2,0}=\{x'|x'\in X_{NS}\cap x_{obs}\}$, and thus $x\in x_{1,0}$ holds.

Let $t=t_1t_2\cdots t_{n}$ such that $|P(t_i)|$ = $1$ for all $i\in \{1,2,\ldots,n\}$.
Thus, for all $i\in \{1,2,\ldots,n\}$, we have $P(t_i)$ = $e_{o,i}$ and $\delta(x,t_1t_2\cdots t_i)\subseteq x_{1,i}\cup x_{2,i}$ due to $x_{st,i}=(x_{1,i},x_{2,i})$.
For all $x_p\in X\cap X_{NS}$ satisfying $x_p\in \delta(x_0',s')$ and $P(s')=P(s)$, it holds that $x_p\in x_{2,0}$ due to the construction of the observer.
For all $t'=t_1't_2'\cdots t_{n}'$ such that $|P(t_i')|=1$ and $P(t_i')=P(t_i)$ for $i= 1,2,\ldots ,n$, the following is true: for all $i\in \{1,2,\ldots ,n\}$, we have $\delta(x_p,t_1't_2'\cdots t_i')\subseteq x_{1,i}\cup x_{2,i}$.
For each $x_{n}'\in x_{2,n}$, we can find an $x_p'\in x_{2,0}$ and a string $t'=t_1't_2'\cdots t_{n}'$ such that $x_{n}\in \delta(x_p',t_1't_1't_2'\cdots t_{n}')$ due to the construction of the state-tree.
Since $x_{2,0}\subseteq X_{NS}$, we have $x_p'\in X_{NS}$.
Therefore, string $s't'$ can be found for string $st$, which contradicts the assumption. Hence, we have $x_{2,n}=\emptyset$, i.e., there exists $x_{st}=(x_1,x_2)\in X_{st}$ such that $x_2$ = $\emptyset$.

($\Longrightarrow$) If $G_{nd}$ is $K$-step weak opaque, then for all $x_{0}\in X_{0}$ and for all $st\in L(G_{nd},x_{0})$ such that $\delta(x_0,s)\cap X_S\neq \emptyset$, $\delta(\delta(x_0,s)\cap X_{S},t)\neq \emptyset$, and $|P(t)|\leq K$, there exist state $x_{0}'\in X_{0}$ and string $s't'\in L(G_{nd},x_{0}')$ satisfying $P(s')=P(s)$ and $P(t')=P(t)$ such that $\delta(x_0',s')\cap X_{NS}\neq \emptyset$ and $\delta(\delta(x_0',s')\cap X_{NS},t')\neq \emptyset$.

We have constructed all state-trees for all $x_{obs}\in X_{obs}$ such that $x_{obs}\cap X_{S}\neq \emptyset$, and thus, we choose an arbitrary $x_{0}\in X_{0}$ and arbitrary $st\in L(G_{nd},x_{0})$ such that $\delta(x_0,s)\cap X_S\neq \emptyset$, $\delta(\delta(x_0,s)\cap X_{S},t)\neq \emptyset$, and $|P(t)|\leq K$.
For $x\in \delta(x_0,s)\cap X_S$ with $\delta(x,t)\neq \emptyset$, we have a state-tree with starting node being $(x_{1,0},x_{2,0})$ and $x\in x_{1,0}$.
A sequence of observations $P(t)[1]P(t)[2]\cdots P(t)[|P(t)|]$ can be found in the state-tree, and for this sequence of observations, the corresponding sequence of nodes is $x_{st,0}x_{st,1}\cdots x_{st,|P(t)|}=(x_{1,0},x_{2,0})(x_{1,1},x_{2,1})\cdots (x_{1,|P(t)|},x_{2,|P(t)|})$.

For all $x_{0}'\in X_{0}$ and $s't'\in L(G_{nd},x_{0}')$ satisfying $P(s')=P(s)$ and $P(t')=P(t)$ such that $x'\in \delta(x_0',s't')$, we have $x'\in x_{1,|P(t)|}\cup x_{2,|P(t)|}$.
Due to the construction of the state-tree, for all $x'\in x_{2,|P(t)|}$, there exist $x''\in x_{2,0}$ and $t'\in E^*$ such that $x'\in \delta(x'',t')$ and $P(t')=P(t)$.
Since $x''\in x_{2,0}$ and $x_{2,0}\subseteq X_{NS}$, for this state-tree corresponding to $x$, we have that for all $i\in \{0,1,\ldots,|P(t)|\}$, $x_{2,i}\neq \emptyset$ holds.
Therefore, for all $x_{st}=(x_1,x_2)\in X_{st}$, we have $x_2\neq \emptyset.$
This completes the proof. $\hfill \square$
\end{pf}

\begin{exmp}
Consider root node $(\{7\},\{4\})$ in Example~\ref{eg-k-step-weak-opaque-yes}. Following Step 2, $((\{7\},\{4\}),c,(\{5\},\{2\}))$ is added. For the new node (\{5\},\{2\}), we obtain the transition $((\{5\},\{2\}),b,(\{6\},\{3\}))$.
The secret state~7 cannot be inferred by an intruder when it tracks back~2 steps.
All state-trees have been constructed in Fig.~\ref{eg4}(c).
According to Theorem~\ref{$K$-step-weak-opaque-verification}, the system is 2-step weak opaque.
\hfill $\square$
\end{exmp}

\begin{figure*}[!htbp]
  \centering
  \includegraphics[width=0.75\linewidth]{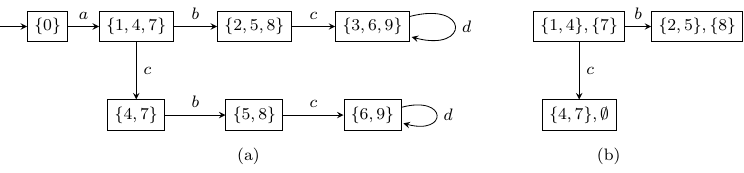}\\
  \caption{Observer and state-tree for the NFA in Fig.~\ref{running-example-2}.}\label{running-example-2-obsever}
\end{figure*}

\textbf{Complexity Analysis.}
The complexity of using Theorem~\ref{$K$-step-weak-opaque-verification} to check $K$-step weak opacity is analysed as follows.
The observer of a considered NFA $G_{nd}=(X,E,\delta,X_0)$ (with respect to a set of observable events $E_o$, a set of secret states $X_S$, and natural projection $P$) has at most $2^{|X|}$ states and $2^{|X|}|E_o|$ transitions. The complexity of constructing the observer is $O(2^{|X|}|E_o|)$.
Since for each $x_{obs}\in X_{obs}$ such that $x_{obs}\cap X_{S}\neq \emptyset$ and $x_{obs}\cap X_{NS}\neq \emptyset$ (if there exists $x_{obs}\in X_{obs}$ such that for all $x\in x_{obs}$, $x\in X_S$, the considered NFA is not current-state opaque, and thus it is not $K$-step weak opaque), the state-tree needs to be computed; note that there are at most $2^{|X|}-2^{|X_{NS}|}-2^{|X_S|}$ state estimates in the observer, for which we need to construct state-trees.
For each $x_{obs}\in X_{obs}$ that we need to construct a state-tree, there are at most $|E_o|$ subsequent nodes (this follows the transition structure of the observer), and for each new state (in the observer), there are also at most $|E_o|$ subsequent nodes.
Since this process needs to be repeated $K-1$ times, the number of nodes in the constructed state-tree for a given root node is at most $1+|E_o|+|E_o|^2+\cdots+|E_o|^K$, and the number of transitions is at most $(|E_o|+|E_o|^2+\cdots+|E_o|^K)|E_o|$.
Therefore, the complexity of constructing all state-trees and verifying the condition in Theorem~\ref{$K$-step-weak-opaque-verification} is $O(|E_o|^{K+1}(2^{|X|}-2^{|X_{NS}|}-2^{|X_S|}))$.
Hence, the complexity of verifying $K$-step weak opacity is $O(\max\{|E_o|^{K+1}(2^{|X|}-2^{|X_{NS}|}-2^{|X_S|}),2^{|X|}|E_o|\})$.

\begin{rem}
The authors of \cite{Xiang:2017} develop an \mbox{algorithm} for checking $K$-step weak opacity with complexity of $O(\min\{|E_o|2^{2|X|}, |E_o|^{K+1}2^{|X|}\})$, where $X$ and $E_o$ are the sets of states and observable events in the considered automaton, respectively.
In \cite{Balun:2021}, approaches to verify $K$-step weak opacity are presented; the computational complexity is $O((K+1)2^{|X|}(|X|+m|E_o|^2))$, where $m\leq |E_o||X|^2$ is the number of edges in the projected automaton \cite{Hop:1979} of the considered automaton.
Our approach has significant complexity advantages in cases when $|E_o|^{K+1}(2^{|X|}-2^{|X_{NS}|}-2^{|X_S|})\leq 2^{|X|}|E_o|$ based on the above descriptions.

Note that a system is infinite-step weak opaque if and only if it is $(2^{|X|}-2)$-step weak opaque (demonstrated in \cite{Xiang:2017}). Thus, the complexity of verifying infinite-step weak opacity using the proposed approach (with $K=2^{|X|}-2$) is $O(\max\{|E_o|^{K+1}(2^{|X|}-2^{|X_{NS}|}-2^{|X_S|}),2^{|X|}|E_o|\})$ = $O(\max\{|E_o|^{(2^{|X|}-1)}(2^{|X|}-2^{|X_{NS}|}-2^{|X_S|}),2^{|X|}|E_o|\})$, which is higher than the existing methods. However, we can assert that the system is not infinite-step weak opaque as soon as an $x_{st}=(x_1,x_2)\in X_{st}$ with $x_2= \emptyset$ is obtained.
\end{rem}
}

\begin{exmp}
This example is used to show the advantage of using the state-tree approach to verify that an NFA is not infinite-step weak opaque.
Reconsider the NFA in Fig.~\ref{running-example-2} and its observer depicted in Fig.~\ref{running-example-2-obsever}(a).
The observer has six states for which state-tree construction are needed.
For state $\{1,4,7\}$, the root node is $(\{1,4\},\{7\})$.
Based on the transition function of the observer, events $b$ and $c$ might be observed at $\{1,4,7\}$;
according to transition function of the NFA, we have $(\{2,5\},\{8\})$ via event $b$ and $(\{4,7\},\emptyset)$ via event $c$, as shown in Fig.~\ref{running-example-2-obsever}(b).
There exists an $x_{st}=(x_{1},x_{2})=(\{4,7\},\emptyset)$ such that $x_{2}=\emptyset$ for the state-tree constructed for state estimate $\{1,4,7\}$, leading to the conclusion that the system is not infinite-step weak opaque (without constructing other state-trees).
\hfill $\square$
\end{exmp}

{\section{Verification for $K$-Step Strong Opacity}
This section presents a method for checking $K$-step strong opacity.
For an NFA, we first compute the secret-involved projected automaton and the observer, based on which $K$-step strong opacity can be checked by constructing secret-unvisited state-trees (SSTs) in a similar manner as verifying $K$-step weak opacity.

Before constructing a secret-involved automaton, the method of obtaining projected automaton in \cite{{Hop:1979}} are recalled as follows. Given an NFA $G_{nd}=(X,E,\delta,X_0)$ with $E=E_{o}\dot{\cup}E_{uo}$, the projected automaton is given by an NFA $G_{p}=(X_{p},E_{o},\delta_{p},X_{0,p})$ with $X_{p}=X$ and $X_{0,p}=UR(X_{0})$, where $\delta_{p}$ is defined as: for $x\in X$ and $e_{o}\in E_{o}$, $\delta_{p}(x, e_{o})= \{x'|(\exists s\in E^{*})[x'\in R(x, s)\wedge e_{o}=P(s)\}$.

Considering secret states of an NFA and the way of constructing its projected automaton, we make the following observations.
Given an NFA $G_{nd}=(X,E,\delta,X_0)$, we assign a subscript $Y$ or $N$ to each state $x\in X$ to indicate whether it is certain that we have reached state $x$ by definitely having visited secret states or whether we are not certain that secret states have been visited: when the NFA reaches $x'$ from any state $x\in X$ via an observable event $e_{o}$ which corresponds to one or more string $s\in E^{*}$ (i.e., such that $P(s)=e_{o}$), $x'_{Y}$ denotes that some secret states definitely have been visited, and $x'_{N}$ denotes that it is possible that no secret state has been visited; we have $X_{N}=\{x_{N}|x\in X\}$ and $X_{Y}=\{x_{Y}|x\in X\}$. Based on the aforementioned descriptions, the following two conditions are provided.

Condition 1: for each $x\in R(X_{0},\varepsilon)$, we have $x_{N}$ if there exist $x_{0}\in X_{0}\cap X_{NS}$ and $s\in L(G_{nd},x_{0})$ satisfying (i) $P(s)=\varepsilon$ and (ii) $x\in \delta(\cdots\delta(\delta(x_{0},s[1])\cap X_{NS},s[2])\cap X_{NS}\cdots, s[|s|])\cap X_{NS}$; otherwise, we have $x_{Y}$.

Condition 2: for all states $x_{N},x_{Y}\in X_{N}\cup X_{Y}$, for all observable events $e_{o}\in E_{o}$ with $e_{o}\in T_o(x)$, and for all $x'\in R(x,e_{o})$, there are two cases: (i) $x\in X_{NS}$ and (ii) $x\in X_{S}$.
For case (i): we have $(x_{N},e_{o},x'_{N})$ and $(x_{Y},e_{o},x'_{N})$ if $x'$ is reached from $x$ via observable event $e_{o}$ under the condition that there exists a string $s\in E^{*}$ with $P(s)=e_{o}$ such that $x'\in \delta(\cdots\delta(\delta(x,s[1])\cap X_{NS},s[2])\cap X_{NS}\cdots, s[|s|])\cap X_{NS}$; otherwise, we have $(x_{N},e_{o},x'_{Y})$ and $(x_{Y},e_{o},x'_{Y})$.
For case (ii): we have $(x_{Y},e_{o},x'_{Y})$.

With Conditions 1 and 2, we are ready to obtain a secret-involved projected automaton for the considered NFA, denoted by $G_{SP}=(X_{SP},E_{SP},\delta_{SP},X_{SP,0})$, where $X_{SP}=\{x_{N}\cup x_{Y}|x\in X\}$, $E_{SP}=E_{o}$, $X_{SP,0}=\{x_{N}\cup x_{Y}|x\in R(X_{0},\varepsilon)\wedge x\text{ satisfies Condition 1}\}$, and $\delta_{SP}$ is defined following Condition 2.
Note that the secret-involved projected automaton is actually an NFA with all events being observable.

\begin{exmp}\label{eg-secret-involved-projected-automaton}
Consider the NFA in Fig.~\ref{k-strong-opaque-NFA}(a) with respect to the set of observable events $E_{o}=\{a,b,c\}$, initial state $X_{0}=\{0\}$, and set of secret states (graphically represented by shaded circles) $X_{S}=\{1\}$.
The initial state 0 is not secret, and $1\in R(0,\varepsilon)$ is secret, leading to $X_{SP,0}=\{0_{N}, 1_{Y}\}$.
It is obvious that $T_{o}(0)=\{a\}$ and $R(0,a)=\{2,4\}$.
The secret state 1 is not necessarily visited when state 2 is reached via the observation $a$ at state 0 since the NFA has transition $(0,a,2)$, and thus we have the transition $(0_{N},a,2_{N})$ in the obtained secret-involved projected automaton as shown in Fig.~\ref{k-strong-opaque-NFA}(b).
Note that since state 2 can be reached via observable event $a$ from secret state 1, we have $(1_{Y},a,2_{Y})$. For $2_{N}$ and $2_{Y}$, transitions $(2_{N},b,3_{N})$ and $(2_{Y},b,3_{N})$ are obtained since state~3 is reached via the observable event $b$ from 2 under the condition that no secret state has been visited.
\hfill $\square$
\end{exmp}

\begin{figure}[!htbp]
  \centering
  \includegraphics[width=\linewidth]{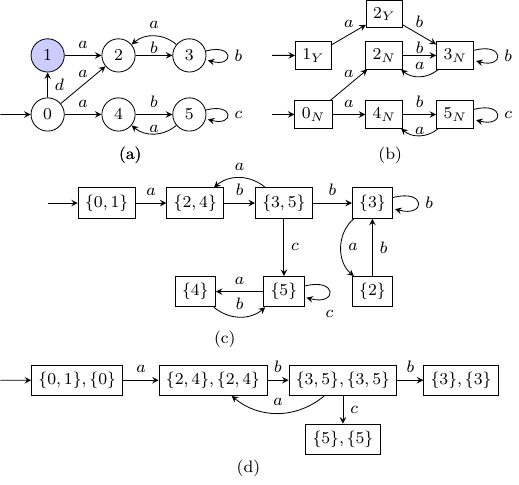}\\
  \caption{Example for secret-involved projected automaton for 3-step strong opacity: (a) NFA with $X_{S}=\{1\}$; (b) Secret-involved projected automaton; (c) Observer; (d) Secret-unvisited state-tree associated with $\{0,1\}$.}\label{k-strong-opaque-NFA}
\end{figure}

Consider each state $x_{obs}\in X_{obs}$ satisfying $x_{obs}\cap X_{S}\neq \emptyset$ as a root node, and then construct a SST for $x_{obs}$ based on the following descriptions.

Step 1: Obtain the root node by defining $rn=(x_1,x_2)$ with $x_{1}=x_{obs}$ and $x_{2}=\{x|x\in x_{obs}\cap X_{NS}\wedge x_{N}\in X_{SP}\}$.

Step 2: Based on the observer, for all $e_{o}\in E_{o}$ with $x_{obs}'= f_{obs}(x_{obs},e_{o})$, the immediate subsequent node corresponding to $x_{obs}'$ in the constructed secret-unvisited state-tree is obtained as $rn'=(x_1',x_2')$ with $x_{1}'= x_{obs}'$ and $x_{2}'=\{x\in x_{obs}'|(\exists x'\in x_2)x\in R(x',e_o)\wedge (x_{N}',e_o,x_{N})\in \delta_{SP}\}$, i.e., the transition $(rn,e_o,rn')$ is added to the SST.

Step 3: Let $rn=rn'$ and repeat Step 3 $(K-2)$ times.

\begin{exmp}\label{eg-k-step-strong-opaque-no}
Reconsider the NFA in Fig.~\ref{k-strong-opaque-NFA}(a) and its observer is attained in Fig.~\ref{k-strong-opaque-NFA}(c). Suppose that we would like to verify 3-step strong opacity. Only one secret-unvisited state-tree needs to be constructed with $\{0,1\}$ being the considered state due to $X_S=\{1\}$.
The root node is obtained as $(\{0,1\},\{0\})$ based on Step 1.
Following Step~2, the transition $((\{0,1\},\{0\}),a,(\{2,4\},\{2,4\}))$ is added into the SST since $T_{o}(\{0,1\})=\{a\}$, $f_{obs}(\{0,1\},a)=\{2,4\}$, $(0_N,a,2_{N})\in \delta_{SP}$, and $(0_N,a,4_{N})\in \delta_{SP}$.
For the new node (\{2,4\},\{2,4\}), we add the transition $((\{2,4\},\{2,4\}),b,(\{3,5\},\{3,5\}))$ into the SST since $T_{o}(\{2,4\})=\{b\}$, $f_{obs}(\{2,4\},b)=\{3,5\}$, $(2_N,b,3_{N})\in \delta_{SP}$, and $(4_N,b,5_{N})\in \delta_{SP}$.
For the new node $(\{3,5\},\{3,5\})$, we have $T_{o}(\{3,5\})=\{a,b,c\}$.
Take event $b$ as an instance. The transition $((\{3,5\},\{3,5\}),b,(\{3\},\{3\}))$ is added into the SST due to $(3_N,b,3_{N})\in \delta_{SP}$.
Following the same procedure, the SST corresponding to $\{0,1\}$ is shown in Fig.~\ref{k-strong-opaque-NFA}(d).
An intruder will never be certain that secret state 1 has been visited when it observes three more events.
Therefore, the NFA is 3-step strong opaque.
\hfill $\square$
\end{exmp}

\begin{prop}\label{observer-SST}
Given an NFA $G_{nd}=(X,E,\delta,X_0)$ with respect to the set of observable events $E_{o}\subseteq E$, the set of secret states $X_S\subseteq X$, and natural projection $P$, its observer is a DFA $G_{obs}=(X_{obs},E_o,f_{obs},x_{obs,0})$ and its secret-involved projected automaton is an NFA $G_{SP}=(X_{SP},E_{SP},\delta_{SP},X_{SP,0})$.
Given $K$, construct a SST for $x_{obs}\in X_{obs}$ with $x_{obs}\cap X_{S}\neq \emptyset$, denoted by $G_{st}(x_{obs})=(X_{st},E_o,f_{st},x_{st,0})$, where $x_{st,0}=(x_{st,0}^{1},x_{st,0}^{2})$ with $x_{st,0}^{1}=x_{obs}$ and $x_{st,0}^{2}=\{x|x\in x_{obs}\cap X_{NS}\wedge x_{N}\in X_{SP}\}$.
Consider a path $p=x_{st,0}x_{st,1}x_{st,2}\cdots x_{st,n}$ in $G_{st}$ and its corresponding sequence of observations $e_{o,1}e_{o,2}\cdots e_{o,n}$ with $n\in\{0,1,\ldots ,K\}$. The following holds:
$$(\forall i\in \{0,1,\ldots,n\})[x_{st,i}^{2}\neq \emptyset]\implies$$
$(\forall x_i\in x_{st,i}^{2})(\exists x_0\in X_{NS}\cap x_{st,0}^{2},$ $\exists s=s_1s_2\ldots s_{i}\in E^{*}$ \text{with} $\forall j\in\{1,2,\ldots,i\} [P(s_j)=e_{o,j}])$ $[x_i\in \delta(\cdots\delta(\delta(x_{0},s[1])\cap X_{NS},s[2])\cap X_{NS}\cdots, s[|s|])\cap X_{NS}$]~\text{and}
$$(\exists i\in \{0,1,\ldots,n\}) [x_{st,i}^{2}=\emptyset]\implies$$
$$(\forall j\in \{i,i+1,\ldots,n\})[x_{st,j}^{2}=\emptyset].$$
\end{prop}

\begin{pf}
\rm
This proof follows immediately from the construction of the observer and the SST.
\end{pf}

\begin{thm}\label{$K$-step-strong-opaque-verification}
Given an NFA $G_{nd}=(X,E,\delta,X_0)$ with set of secret states $X_S\subseteq X$, set of observable events $E_{o}\subseteq E$, and natural projection $P$, and its observer $G_{obs}=(X_{obs},E_o,f_{obs},x_{obs,0})$, construct all SSTs for all $x_{obs}\in X_{obs}$ such that $x_{obs}\cap X_S\neq \emptyset$ and use $X_{st}^{G_{nd}}$ to denote the set of all states in the obtained SSTs. The automaton $G_{nd}$ is $K$-step strong opaque if and only if
$$(\forall x_{st}=(x_{st}^{1},x_{st}^{2})\in X_{st}^{G_{nd}})x_{st}^{2}\neq \emptyset.$$
\end{thm}

\begin{pf}
\rm
($\Longleftarrow$) Suppose that $G_{nd}$ is not $K$-step strong opaque.
Our goal is to find an $x_{st}=(x_{st}^{1},x_{st}^{2})\in X_{st}^{G_{nd}}$ such that $x_{st}^{2}=\emptyset$.
Since $G_{nd}$ is not $K$-step strong opaque, there necessarily exist $x_{0}\in X_{0}$ and $st\in L(G_{nd},x_{0})$ such that $\delta(x_0,s)\cap X_{S}\neq \emptyset$, $\delta(\delta(x_0,s)\cap X_{S},t)\neq \emptyset$, and $|P(t)|\leq K$, and for this string $st$, we conclude that for all $x_{0}'\in X_{0}$ and for all $\omega\in L(G_{nd},x_{0}')$ with $P(\omega)=P(st)$, i.e., for all $\omega\in L(G_{nd})$ with $P(\omega)=P(st)$, the following holds: there exists $\omega'\in \bar{\omega}$ such that $|P(\omega)|-|P(\omega')|\leq K$ and $\delta(\cdots\delta(\delta(\delta(x_0,\omega')\cap X_{NS},\omega''[1])\cap X_{NS},\omega''[2])\cap X_{NS}$$ $$\cdots,\omega''[|\omega''|])\cap X_{NS}=\emptyset$ with $\omega''=\omega/\omega'$.

Let $X_{1}=\delta(x_0,s)\cap X_S$ with $\delta(X_{1},t)\neq \emptyset$, $|P(t)|=n\leq K$, and $P(t)=e_{o,1}e_{o,2}\cdots e_{o,n}$.
We use $x_{st,0}x_{st,1}\cdots x_{st,n}$ to denote the path in the constructed SSTs with the following properties: (i) for all $i\in \{1,2,\ldots,n\}$, we have the transition $(x_{st,i-1},e_{oi},x_{st,i})$, and
(ii) we have $x_{st,0}=(x_{st,0}^{1},x_{st,0}^{2})$ with $x_{st,0}^{1}=x_{obs}=f_{obs}(x_{obs,0},P(s))$ and $x_{st,0}^{2}=\{x'|x'\in X_{NS}\cap x_{obs}\wedge x'_{N}\in X_{SP}\}$, leading to $X_{1}\subseteq x_{st,0}^{1}$.

For all $s'\in L(G_{nd})$ with $P(s')$ = $P(s)$, due to $f_{obs}(x_{obs,0},P(s')) = f_{obs}(x_{obs,0},P(s)) = x_{obs}$ (according to the construction of observer), we have $\delta(X_0,s')\subseteq x_{obs}$. Hence, for all $x\in \delta(X_0,s')\cap X_{NS}$ with $x_{N}\in X_{SP}$, $x\in x_{st,0}^{2}$ holds since $x_{st,0}^{2}=\{x'|x'\in x_{obs}\cap X_{NS}\wedge x'_{N}\in X_{SP}\}$.
For all $t'$ with (i) $s't'\in L(G_{nd})$, (ii) $P(s't')=P(st)$, and (iii) for all $t''\in \bar{t'}$ such that $\delta(\cdots\delta(\delta(\delta(x_0,t'')\cap X_{NS},\sigma[1])\cap X_{NS},\sigma[2])\cap X_{NS}$$ $$\cdots,\sigma[|\sigma|])\cap X_{NS} \neq \emptyset$ with $\sigma=t'/t''$, we have $\delta(\delta(X_0,s'),t')\subseteq x_{st,n}^{2}$ due to the construction of the SST and the secret-involved projected automaton.
According to Proposition~\ref{observer-SST}, for each $x_{n}\in x_{st,n}^{2}$, we can find an $x_p\in x_{st,0}^{2}$ and a string $t'$ such that (i) $s't'\in L(G_{nd})$, (ii) $P(s't')=P(st)$, and (iii) $\delta(\cdots\delta(\delta(\delta(X_{0},s')\cap X_{NS},t'[1])\cap X_{NS},t'[2])\cap X_{NS}\cdots,t'[|t'|])\cap X_{NS}\neq \emptyset$.
Let $\omega=s't'$.
For all $\omega'\in \bar{\omega}$ such that $|P(\omega)|-|P(\omega')|\leq n$ and $\delta(\cdots\delta(\delta(\delta(x_0,\omega')\cap X_{NS},\omega''[1])\cap X_{NS},\omega''[2])\cap X_{NS}$$ $$\cdots,\omega''[|\omega''|])\cap X_{NS}\neq \emptyset$ with $\omega''=\omega/\omega'$, we have $\delta(X_0,\omega')\cap X_{NS}\subseteq x_{st,i}^{2}$ if $|P(\omega)|-|P(\omega')|=i$ with $i\in \{0,1,\ldots,n\}$, which contradicts the assumption.
Therefore, there exists $j\in \{0,1,\ldots,n\}$ such that $x_{st,j}^{2}=\emptyset$.

($\Longrightarrow$) If $G_{nd}$ is $K$-step strong opaque, then for all $x_{0}\in X_{0}$ and for all $st\in L(G_{nd},x_{0})$ such that $\delta(x_0,s)\cap X_S\neq \emptyset$, $\delta(\delta(x_0,s)\cap X_{S},t)\neq \emptyset$, and $|P(t)|\leq K$, there exist $x_{0}'$ and $\omega\in L(G_{nd},x_{0}')$ satisfying the following two conditions: (i) $P(\omega)=P(st)$ and (ii) for all $\omega'\in \bar{\omega}$ with $P|(\omega)|-|P(\omega')|\leq K$, $\delta(\cdots\delta(\delta(\delta(x_0,\omega')\cap X_{NS},\omega''[1])\cap X_{NS},\omega''[2])\cap X_{NS} \cdots,\omega''[|\omega''|])\cap X_{NS}\neq \emptyset$ holds, where $\omega''=\omega/\omega'$.
We choose arbitrary $x_{0}\in X_{0}$ and arbitrary $st\in L(G_{nd},x_{0})$ such that $\delta(x_0,s)\cap X_S\neq \emptyset$, $\delta(\delta(x_0,s)\cap X_{S},t)\neq \emptyset$, and $|P(t)|=n\leq K$.
Suppose $P(t)=e_{o,1}e_{o,2}\cdots e_{o,n}$.
For this sequence of observations, the corresponding sequence of nodes is unique (due to the construction of the observer) and can be found in the constructed SSTs with starting node being $(x_{st,0}^{1},x_{st,0}^{2})$ and $(\delta(x_0,s)\cap X_S)\subseteq x_{st,0}^{1}$.
The sequence of nodes is denoted by $x_{st,0}x_{st,1}\cdots x_{st,n} = (x_{st,0}^{1},x_{st,0}^{2})(x_{st,1}^{1},x_{st,1}^{2})\cdots (x_{st,n}^{1},x_{st,n}^{2})$.
For all $i\in \{1,2,\\\ldots,n\}$ with $n\leq K$, $x_{st,i}$ = $f_{st}(x_{st,i-1},e_{oi})$ holds.
For all strings $\omega\in L(G_{nd})$ satisfying $P(\omega)=P(st)$, we have $\delta(X_0,\omega)=x_{st,n}^{1}$ due to the construction of the observer.

Based on Proposition~\ref{observer-SST}, if for all $i\in \{0,1,2,\ldots,n\}$, $x{_{st,i}^{2}}\neq \emptyset$, then for each $x_{i}\in x{_{st,i}^{2}}$, there exist $x_0'\in X_{NS}\cap x_{st,0}^{2}$ and $s' = s_1's_2'\ldots s_{i}'\in E^{*}$ with $P(s_j')=e_{o,j}$ for $j=1,2,\ldots,i$ such that $x_i\in \delta(\cdots\delta(\delta(x_{0}',s'[1])\cap X_{NS},s'[2])\cap X_{NS}\cdots, s'[|s|])\cap X_{NS}$.
Due to $x_{0}'\in x{_{st,0}^{2}}\subseteq x_{st,0}^{1}$, there exists $s''$ with $P(s'')=P(s)$ such that $x_{0}'\in \delta(X_{0},s'')$.
Let $\omega=s''s'$. Thanks to $P(s'')=P(s)$, $P(s')=P(t)$ and $|P(s')|=n<=K$, for all $\omega''\in \bar{\omega}$ such that $|P(\omega)|-|P(\omega'')|\leq K$, we have $\omega''\in \bar{s'}$.
Therefore, a string $\omega$ is found so as to confuse the intruder that the secret states $\delta(x_{0},s)\cap X_{S}$ may have not been visited with $n$ ($n\leq K$) steps being $e_{o,1}e_{o,2}\cdots e_{o,n}$ thereafter.
SSTs are constructed for all $x_{obs}\in X_{obs}$ such that $x_{obs}\cap X_S\neq \emptyset$.
Hence, for all $x_{st}=(x_{st}^{1},x_{st}^{2})\in X_{st}$, we have $x_{st}^{2}\neq \emptyset.$
This completes the proof. \hfill$\square$
\end{pf}

\textbf{Complexity Analysis.}
The complexity of using Theorem~\ref{$K$-step-strong-opaque-verification} to check $K$-step strong opacity is analysed as follows.
The observer of a considered NFA $G_{nd}=(X,E,\delta,X_0)$ (with respect to set of observable events $E_o$, set of secret states $X_S$, and natural projection $P$) has at most $2^{|X|}$ states and $2^{|X|}|E_o|$ transitions. The complexity of constructing the observer is $O(2^{|X|}|E_o|)$.
The secret-involved projected automaton of $G_{nd}$ is an NFA and the number of states and transitions are at most $2|X|$ and $(2|X|)^2|E_o|=4|X|^2|E_o|$, respectively. The complexity of constructing the secret-involved projected automaton is $O(|X|^2|E_o|)$.
Since for each $x_{obs}\in X_{obs}$ such that $x_{obs}\cap X_{S}\neq \emptyset\wedge x_{obs}\cap X_{NS}\neq \emptyset$ (if there exists $x_{obs}\in X_{obs}$ such that for all $x\in x_{obs}$, $x\in X_S$ holds, the considered NFA is not current-state opaque and thus it is not $K$-step strong opaque), the SST needs to be computed, and thus there are at most $2^{|X|}-2^{|X_{NS}|}-2^{|X_S|}$ state estimates in the observer, for which we need to construct SSTs.
For each $x_{obs}\in X_{obs}$ that needs a SST construction, there are at most $|E_o|$ subsequent nodes (this follows the transition structure of the observer), and for each new state (in the observer), there are also at most $|E_o|$ subsequent nodes.
Since this process needs to be repeated $(K-1)$ times, the number of nodes in the constructed SST for a given root node is at most $1+|E_o|+|E_o|^2+\cdots+|E_o|^K$, and the number of transitions is at most $(|E_o|+|E_o|^2+\cdots+|E_o|^K)|E_o|$.
Therefore, the complexity of constructing all SSTs and verifying the condition in Theorem~\ref{$K$-step-strong-opaque-verification} is $O(|E_o|^{K+1}(2^{|X|}-2^{|X_{NS}|}-2^{|X_S|}))$.
Hence, the complexity of verifying $K$-step strong opacity is $O(\max\{|E_o|^{K+1}(2^{|X|}-2^{|X_{NS}|}-2^{|X_S|}),2^{|X|}|E_o|\})$.

\begin{rem}
In \cite{Ma:2021}, a $K$-step recognizer is constructed to check $K$-step strong opacity for a given value of $K$, with complexity $O(2^{(K+2)|X|}|E_o|)$.
Our method has a clear complexity advantage in verifying $K$-step strong opacity and is also applicable to NFAs.
\end{rem}

\begin{exmp}\label{eg-k-step-strong-opaque-yes}
Consider the DFA in Fig.~\ref{k-strong-opaque}(a) with respect to the set of observable events $E_{o}=\{a,b,c\}$, initial state $x_{0}=0$, and set of secret states (graphically represented by shaded circles) $X_{S}=\{1\}$.
The obtained secret-involved projected automaton, the observer, and the SST corresponding to $\{0,1\}$ are shown in Fig.~\ref{k-strong-opaque}(b), (c), and (d), respectively.
Based on Theorem~\ref{$K$-step-strong-opaque-verification}, the DFA is not 3-step strong opaque. When an intruder observes $abb$, it is certain that secret state 1 has been visited over the last three steps.
\hfill $\square$
\end{exmp}

\begin{figure}[!htbp]
  \centering
  \includegraphics[width=0.98\linewidth]{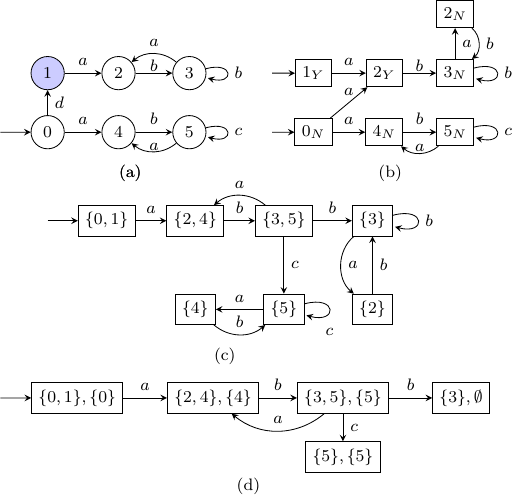}\\
  \caption{Example DFA that is not 3-step strong opaque: (a) DFA with $X_{S}=\{1\}$; (b) Secret-involved projected automaton; (c) Observer; (d) Secret-unvisited state-tree associated with $\{0,1\}$.}\label{k-strong-opaque}
\end{figure}

}
\begin{exmp}
Reconsidering the NFA in Fig.~\ref{running-example}, we have its secret-involved projected automaton and observer  as depicted in Fig.~\ref{running-example-obsever}(a) and (b), respectively.
The observer has three states for which we need to obtain SSTs.
Suppose that we would like to verify 2-step strong opacity.
For state $\{1,4,7\}$, the root node is $(\{1,4\},\{7\})$.
Based on the transition function of the observer, event $b$ might be observed at $\{1,4,7\}$.
According to transition function of the NFA, we have $(\{2,5,8\},\{8\})$ via event $b$ at $(\{1,4\},\{7\})$, as shown on the top of Fig.~\ref{running-example-obsever}(c).
All SSTs are obtained in Fig.~\ref{running-example-obsever}(c).
There exists an $x_{st}=(x_{1},x_{2})=(\{3,6,9\},\emptyset)$ such that $x_{2}=\emptyset$ for the SST constructed for state estimate $\{1,4,7\}$.
According to Theorem~\ref{$K$-step-strong-opaque-verification}, the system is not 2-step strong opaque, which matches with Example~\ref{eg-1}.
Note that, when verifying $K$-step strong opacity using the method of constructing SSTs, as soon as an $x_{st}=(x_{1},x_{2})$ with $x_{2}=\emptyset$ is obtained, the system is not $K$-step strong opaque.
\hfill $\square$
\end{exmp}

\begin{figure}[!htbp]
  \centering
  \includegraphics[width=0.88\linewidth]{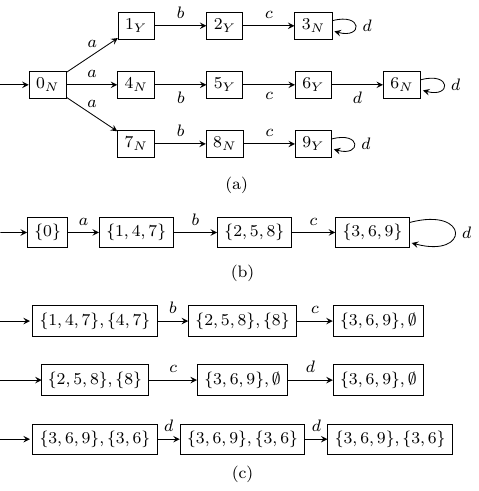}\\
  \caption{Example for $K$-step strong opacity.}\label{running-example-obsever}
\end{figure}

{\section{Verification for Infinite-Step Strong Opacity}
This section focuses on the verification for infinite-step strong opacity.
Given a partially-observable NFA $G_{nd}=(X,E,\delta,X_0)$, we first construct the secret-involved projected automaton and the observer, based on which a verifier is obtained to check infinite-step strong opacity.

%
%
%
%

Algorithm~\ref{alg:verifier-strong} below is presented to construct a verifier based on the secret-involved projected automaton and the observer for checking infinite-step strong opacity.
The algorithm takes as inputs an NFA, its observer and secret-involved projected automaton. The obtained verifier is initialized with the initial state; the initial state of the verifier is a pair of state set, which consists of the initial state of the observer and the system states in the initial state of the observer that correspond to initial states of the secret-involved projected automaton with a subscript $N$. The algorithm iterates over each state in the verifier and computes the observable events based on the first element of the verifier state (i.e., the state of the observer). For each observable event, it calculates the next verifier state by applying the observer's transition function and filtering the resulting states based on the transition function of the secret-involved projected automaton. We add the transition from the current examined verifier state to the new obtained verifier state via the observable event to the verifier.
If the new obtained state does not belong to the verifier, then add the state to the verifier. This process continues until all states in the verifier have been examined, resulting in the final verifier that captures all sequence of observations of the original NFA while incorporating the necessary secrecy constraints.

\begin{algorithm}[!htbp]
        \caption{Construction of a verifier based on observer and secret-involved projected automaton}\label{alg:verifier-strong}
        \begin{algorithmic}[1]
        \Require {NFA $G_{nd}=(X,E,\delta,X_0)$ and its observer $G_{obs}=(X_{obs},E_{o},f_{obs},x_{obs,0})$ and secret-involved projected automaton $G_{SP}=(X_{SP},E_{SP},\delta_{SP},X_{SP,0})$}
        \Ensure  {Verifier $G_{v}=(X_{v},E_{o},f_{v},x_{v,0})$}
        \State Let $x_{v,0}=(x_{obs,0},\{x\in x_{obs,0}|x_N\in X_{SP,0}\})$ and $X_{v}=\{x_{v,0}\}$
        \For{$x_{v}\in X_{v}$ that has not been examined}
        \State Compute $T_o(x{_{v}^{1}})$ /* We use $x{_{v}^{1}}$ and $x{_{v}^{2}}$ to respectively denote the first and second element in $x_{v}$.*/
        \For{$e_{o}\in T_o(x{_{v}^{1}})$ that has not been examined}
            \State Find $x{_{v}^{1}}'=f_{obs}(x{_{v}^{1}},e_{o})$ and $x{_{v}^{2}}'=\{x'\in x{_{v}^{1}}'|\exists x\in x{_{v}^{2}}[(x_{N},e_{o},x_N')\in \delta_{SP}]\}$
            \State Obtain $x_{v}'=(x{_{v}^{1}}',x{_{v}^{2}}')$
            \State Add transition $(x_{v},e_{o},x_{v}')$ into $f_{v}$
            \If{$x_{v}'\notin X_{v}$}
                \State Add $x_{v}'$ into $X_{v}$
            \EndIf
            \State Mark that $e_{o}$ is examined
            \EndFor
        \State Mark that $x_{v}\in X_{v}$ is examined
        \EndFor
\end{algorithmic}
\end{algorithm}

\begin{exmp}
Consider the DFA in Fig.~\ref{DFA-verifier-infinite-step-opacity}(a) with \mbox{respect} to the set of observable events $E_{o}$ = $\{a,c\}$, the set of unobservable events $\{b\}$, and the set of secret states (graphically represented by shaded circles) $X_S$ = $\{3,6\}$.
The observer and the secret-involved projected automaton are constructed as shown in Fig.~\ref{DFA-verifier-infinite-step-opacity}(b) and Fig.~\ref{DFA-verifier-infinite-step-opacity}(c), respectively.
We use this example to show the process of constructing a verifier based on Algorithm~\ref{alg:verifier-strong}.
In the observer, we have $x_{obs,0}=\{0,1\}$, and in the secret-involved projected automaton, we have $0_{N},1_{N}\in X_{SP,0}$. Therefore, in line~1, the following holds: $x_{v,0} = (\{0,1\},\{0,1)\})$ and $X_v=\{x_{v,0}\}$.
In lines 2 and 3, for state $(\{0,1\},\{0,1\})$, the set of events that can occur at $\{0,1\}$ in the observer is $T_{o}(\{0,1\})=\{a,c\}$.
In lines 4--7, for $a$, $f_{obs}(\{0,1\},a) = \{2,3,5,6\}$ holds; in $G_{SP}$, there exist $(0_N,a,2_N)$, $(0_N,a,5_N)$, $(1_N,a,2_N)$, and $(1_N,a,5_N)$; but there do not exist $(0_N,a,3_N)$, $(0_N,a,6_N)$, $(1_N,a,3_N)$, or $(1_N,a,6_N)$; therefore, $f_{v}((\{0,1\},\{0,1\}),a) = (\{2,3,5,6\},\{2,5\})$ holds, and we add it to $f_{v}$. Based on lines 8--10, we add $(\{2,3,5,6\},\{2,\\5\})$ to $X_v$, i.e., $X_v = \{(\{0,1\}, \{0,1\}), (\{2,3,5,6\},\{2,5\})\}$.
After repeating the process as aforementioned, the verifier is completed as shown in Fig.~\ref{DFA-verifier-infinite-step-opacity}(d) following Algorithm~\ref{alg:verifier-strong}.
 \hfill $\square$
\end{exmp}

\begin{figure}[!htbp]
\centering
\includegraphics[width=\linewidth]{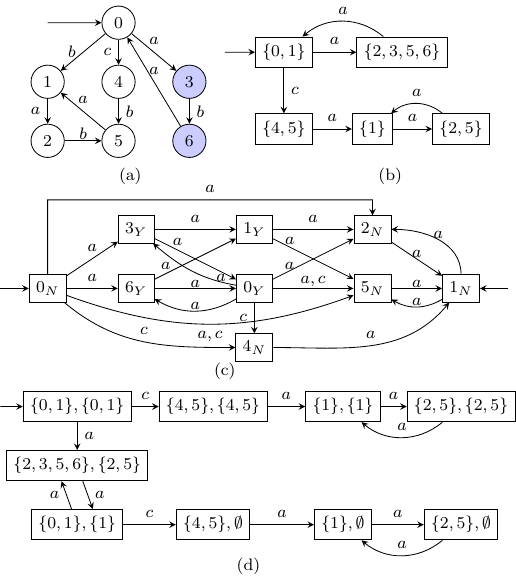}
\caption{Example for infinite-step strong opacity: (a) DFA with $X_{S}=\{3,6\}$; (b) Observer; (c) Secret-involved projected automaton; (d) Verifier.}\label{DFA-verifier-infinite-step-opacity}
\end{figure}

\begin{prop}\label{verifier-and-observer-strong}
Consider an NFA $G_{nd}=(X,E,\delta,X_0)$ and its verifier $G_{v}=(X_{v},E_{o},f_{v},x_{v,0})$ constructed based on Algorithm~\ref{alg:verifier-strong}. Then, $L(G_{v})=L(G_{obs})$ holds.
\end{prop}

\begin{pf}
\rm
Consider an arbitrary sequence of observations $\omega\in L(G_{obs})$.
Since $G_{obs}$ is a fully observable DFA, there necessarily exists a unique sequence of states in the form of $x_{obs,0}x_{obs,1}\cdots x_{obs,|\omega|}$, for which the following holds: for all $i\in \{1,2,\ldots, |\omega|\}$, $f_{obs}(x_{obs,i-1},\omega[i])=x_{obs,i}.$
According to the construction of the verifier $G_{v}$ following Algorithm~\ref{alg:verifier-strong}, there necessarily exists a unique sequence of states in $G_{v}$ in the form of $x_{v,0}x_{v,1}\cdots x_{v,|\omega|}$, for which the followings hold: (i) for all $i\in \{0,1,2,\ldots, |\omega|\}$, $x{_{v,i}^{1}}=x_{obs,i}$, and (ii) for all $i\in \{1,2,\ldots, |\omega|\}$, $f_{v}(x_{v,i-1},\omega[i])=x_{v,i}.$
As a consequence, the sequence of observations that corresponds to $x_{v,0}x_{v,1}\cdots x_{v,|\omega|}$ is $\omega$. Therefore, $\omega\in L(G_{v})$ holds, leading to $L(G_{obs})\subseteq L(G_{v})$.

Considering an arbitrary sequence of observations $\omega\in L(G_{v})$, there necessarily exists a unique sequence of states in $G_{v}$ in the form of $x_{v,0}x_{v,1}\cdots x_{v,|\omega|}$ such that for all $i\in \{1,2,\ldots, |\omega|\}$, $f_{v}(x_{v,i-1},\omega[i])=x_{v,i}.$
According to the construction of the verifier $G_{v}$ following Algorithm~\ref{alg:verifier-strong}, for all $j\in \{0,1,\ldots, |\omega|\}$, we have $x{_{v,j}^{1}}=x_{obs,j}$, and there necessarily exists a unique sequence of states in $G_{obs}$ in the form of $x_{obs,0}x_{obs,1}\cdots x_{obs,|\omega|}$ such that for all $i\in \{1,2,\ldots, |\omega|\}$, $f_{obs}(x_{obs,i-1},\omega[i])=x_{obs,i}.$
Thus, the sequence of observations that corresponds to $x_{obs,0}x_{obs,1}\cdots x_{obs,|\omega|}$ is $\omega$. Therefore, we have $\omega\in L(G_{obs})$ and $L(G_{v})\subseteq L(G_{obs})$. Finally, $L(G_{v})= L(G_{obs})$ holds.
\hfill $\square$
\end{pf}

\begin{thm}\label{infinite-step-strong-opacity-verification}
Consider an NFA $G_{nd}=(X,E,\delta,X_0)$, with set of secret states $X_S\subseteq X$, set of observable events $E_{o}\subseteq E$, and natural projection $P$, and its observer $G_{obs}$ as well as verifier $G_{v}$. NFA $G_{nd}$ is infinite-step strong opaque if and only if
$$(\forall x_{v}\in X_{v})x{_{v}^{2}}\neq \emptyset.$$
\end{thm}

\begin{pf}
\rm
($\Longrightarrow$)
Assume that there exists $x_{v}\in X_{v}$ in $G_{v}$ such that $x{_{v}^{2}}=\emptyset$.
There necessarily exists a sequence of observations $\omega\in L(G_{v})$ such that $f_{v}(x_{v,0},\omega)=x_{v}$.
From the construction of verifier following Algorithm~\ref{alg:verifier-strong} and the construction of the secret-involved projected automaton, for all $i\in \{0,1,\ldots,|\omega|\}$, $x{_{v,i}^{2}}$ collects all states $x\in X$ such that there exist $x_{0}\in X_{0}$ and $s\in L(G_{nd},x_{0})$ with $P(s)=\omega[1]\omega[2]\cdots \omega[i]$ such that $x\in \delta(\cdots\delta(\delta(x_{0},s[1])\cap X_{NS},s[2])\cap X_{NS}\cdots, s[|s|])\cap X_{NS}$.
By $x{_{v,|\omega|}^{2}}=x{_{v}^{2}}=\emptyset$, for all $x_{0}'\in X_{0}$ and $s'\in L(G_{nd},x_{0}')$ with $P(s')=\omega$, there exists $s''\in \overline{s'}$ such that $\delta(x_{0}',s'')\cap X_{NS}=\emptyset$.
Therefore, there exist $x_{0}\in X_{0}$ and $st\in L(G_{nd},x_{0})$ such that $\delta(x_{0},s)\cap X_{S}\neq \emptyset$ and $\delta(\delta(x_0,s)\cap X_{S},t)\neq \emptyset$, and we cannot obtain $x_{0}'\in X_{0}$ and $s_{1}\in L(G_{nd},x_{0}')$ such that $P(s_{1})=P(st)$ and $\delta(\cdots\delta(\delta(x_{0}',s_{1}[1])\cap X_{NS},s_{1}[2])\cap X_{NS}\cdots,s_{1}[|s_{1}|])\cap X_{NS}\neq \emptyset$.
According to Definition~\ref{infinite-step-strong-opacity}, $G_{nd}$ is not infinite-step strong opaque.
The above descriptions mean that if  $G_{nd}$ is infinite-step strong opaque, in the corresponding verifier $G_{v}$ constructed by Algorithm~\ref{alg:verifier-strong}, for all $x_{v}\in G_{v}$, $x{_{v}^{2}}\neq \emptyset$ holds.

($\Longleftarrow$) We show that if for all $x_{v}\in G_{v}$, $x{_{v}^{2}}\neq\emptyset$ holds, system $G_{nd}$ is infinite-step strong opaque.
Consider an arbitrary sequence of states in $G_{v}$ corresponding to the sequence of observations $\omega$, denoted by $x_{v,0}x_{v,1}\cdots x_{v,|\omega|}$.
For this sequence of states, due to the constructions of the observer, the secret-involved projected automaton and the verifier in Algorithm~\ref{alg:verifier-strong}, the following condition \mbox{holds:} for each $x\in x{_{v,|\omega|}^{2}}$, there exist at least one initial state $x_{0}\in X_{0}$ and one string $s\in L(G_{nd},x_{0})$ such that $P(s)=\omega$ and $ \delta(\cdots\delta(\delta(x_{0},s[1])\cap X_{NS},s[2])\cap X_{NS},\cdots, s[|s|])\cap X_{NS}\neq \emptyset$.
Hence, for all strings $st\in L(G_{nd})$ with $\delta(X_0,s)\cap X_S\neq \emptyset$ and $\delta(\delta(X_0,s)\cap X_S,t)\neq \emptyset$, there exists a unique sequence of states $x_{v,0}x_{v,1}\cdots x_{v,|P(s)|}x_{v,|P(s)|+1}\cdots x_{v,|P(s)|+|P(t)|}$ in $G_{v}$ such that $\delta(X_{0},s)$ = $x{_{v,|P(s)|}^{1}}$ and $P(st)$ = $P(s)[1]P(s)[2]\cdots$
$P(s)[|P(s)|]P(t)[1]P(t)[2]\cdots P(t)[|P(t)|]$.
For each $x\in x{_{v,|P(s)|+|P(t)|}^{2}}$, there exist $x_{0}\in X_{0}$ and $s_{1}\in L(G_{nd},x_{0})$ such that $P(s_{1})=P(s)[1]P(s)[2]\cdots P(s)[|P(s)|]P(t)[1]P(t)[2]\\\cdots P(t)[|P(t)|]=P(st)$  and  $\delta(\cdots\delta(\delta(x_{0},s_{1}[1])\cap X_{NS},\\s_{1}[2])\cap X_{NS}\cdots, s_{1}[|s_{1}|])\cap X_{NS}\neq \emptyset$.
By Definition~\ref{infinite-step-strong-opacity}, system $G_{nd}$ is infinite-step strong opaque.
The proof is completed.
\hfill $\square$
\end{pf}

\begin{exmp}
There exist $(\{4,5\},\emptyset)$, $(\{1\},\emptyset)$, and $(\{2,5\},\emptyset)$ in the verifier of Fig.~\ref{DFA-verifier-infinite-step-opacity}(d) such that $x{_{v}^{2}}=\emptyset$, which means that the sequences of observations that reach states $(\{4,5\},\emptyset)$, $(\{1\},\emptyset)$, and $(\{2,5\},\emptyset)$ will allow an intruder to be certain that some secret states have been visited. Thus, infinite-step strong opacity is violated.
\end{exmp}

\textbf{Complexity Analysis.}
The complexity of obtaining the observer $G_{obs}$ for an NFA $G_{nd}=(X,E,\delta,X_0)$ is $O(2^{|X|}|E_{o}|)$, and the complexity of obtaining the secret-involved projected automaton is $O(|X|^{2}|E_{o}|)$.
According to Algorithm~\ref{alg:verifier-strong}, the number of states in the verifier $G_{v}$ for an NFA $G_{nd}=(X,E,\delta,X_0)$ is at most $(2^{|X|})^2 = 2^{2|X|}$, and the number of transitions in $G_{v}$ is at most $2^{2|X|}|E_{o}|$.
The complexity of checking infinite-step strong opacity using Theorem~\ref{infinite-step-strong-opacity-verification} is $O(2^{2|X|})$.
Therefore, the complexity for verifying infinite-step strong opacity of a system in the worst scenario is $O(2^{2|X|}|E_{o}|)$.

\begin{rem}
The complexity in this paper to verify infinite-step strong opacity is identical to the method in \cite{Ma:2021} that focuses on DFAs. Notice, however, that our approach is applicable to NFAs. During the construction of the verifier in Algorithm~\ref{alg:verifier-strong}, we can assert that the system is not infinite-step strong opaque once $x_{v}$=$(x{_{v}^{1}},x{_{v}^{2}})$ with $x{_{v}^{2}}=\emptyset$ is obtained.
\end{rem}

}

\begin{figure}[!htbp]
  \centering
  \includegraphics[width=0.9\linewidth]{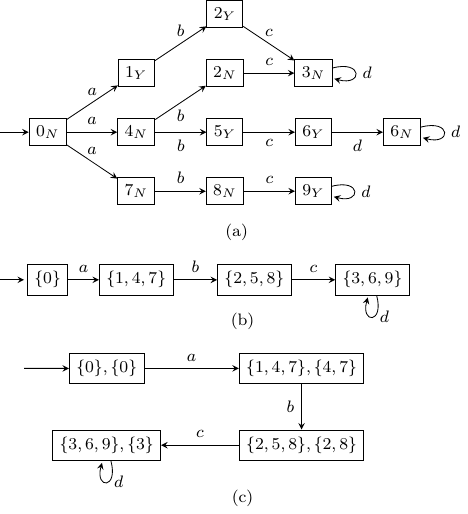}\\
  \caption{Example for infinite-step strong opacity.}\label{running-example-1-obsever}
\end{figure}

\begin{exmp}
Reconsider the NFA in Fig.~\ref{running-example-1}. Its secret-involved projected automaton and observer are obtained as depicted in Fig.~\ref{running-example-1-obsever}(a) and (b), respectively.
According to Algorithm~\ref{alg:verifier-strong}, the corresponding verifier is obtained as shown in Fig.~\ref{running-example-1-obsever}(c).
There does not exist an $x_{st}=(x_{1},x_{2})$ in the verifier such that $x_{2}=\emptyset$.
Based on Theorem~\ref{infinite-step-strong-opacity-verification}, the system is infinite-step strong opaque.
\hfill $\square$
\end{exmp}

{
\section{Conclusions and Future works}
Verifications for $K$-step and infinite-step weak and strong opacity under the framework of nondeterministic finite state automaton with partially unobservable events are addressed in this work.
State-tree, secret-unvisited state-tree, secret-involved projected automaton, and verifier are constructed and utilized to verify the opacity notions formulated in this paper.
In the future, we are interested in enforcing these kinds of opacity under constraints using extended insertion functions in \cite{li2021extended} and \cite{li-2023}.
}


\bibliographystyle{IEEEtran}
\bibliography{IEEEtran}
\end{document}